\documentstyle[preprint,floats,aps,prl] {revtex}

\begin{document}
\draft

\title{Influence of molecular vibrations on dissociative adsorption}

\author{Axel Gross and Matthias Scheffler}

\address{Fritz-Haber-Institut der Max-Planck-Gesellschaft, Faradayweg 4-6, 
D-14195 Berlin-Dahlem, Germany}

\maketitle

\begin{abstract}

The influence of molecular vibrations on dissociative adsorption
is studied by six-dimensional quantum dynamical calculations. For the system
H$_2$ at Pd\,(100), which possesses non-activated pathways,
it is shown that large vibrational effects exist and that they 
are not due to a strongly curved reaction path and a late 
dissociation-hindering minimum barrier, 
as was previously assumed. Instead, they are caused
by the lowering of the H-H vibrational 
frequency during the 
dissociation
and the multi-dimensionality of the 
potential energy surface.
Still there are quantitative discrepancies between theory and experiment
identified.

\end{abstract}


\section{Introduction}

In recent years the dissociative adsorption and associative desorption of 
hydrogen on metal surfaces has been the subject of many experimental and
theoretical investigations (see, e.g., Refs.~\cite{Ren94,Hol94,Dar95} and
references therein). The studies were motivated by the fact that for 
these relatively simple systems the process of bond breaking and bond 
making during dissociative adsorption can be analyzed in greater detail.

Breaking a molecular bond is obviously coupled to the 
interatomic distance and molecular vibrations. Therefore
a large number of studies have addressed the influence of
those vibrations on the dissociative adsorption and associative desorption, 
especially in the benchmark system
H$_2$/Cu~\cite{Kub85,Ret93,Ang89,Hay89,Ret92,Ret95}.
In desorption 
studies~\cite{Kub85,Ret93} strong vibrational heating of the hydrogen 
molecules was found, i.e., vibrational population ratios were
much greater than expected for thermal equilibrium with the temperature 
of the substrate at
which desorption occurs. According to the principle of microscopic
reversibility this implies that the probability of dissociative adsorption
should be enhanced for vibrational excited molecules, which indeed has been
confirmed experimentally~\cite{Ang89,Hay89,Ret92,Ret95}.
These vibrational effects 
were usually discussed~\cite{Jac87,Hal90,Kue91,Dar92a}
within the context of 
potential energy surfaces (PESs) which depend only on two
coordinates, namely the center of mass distance of the molecule from the
surface, $Z$, and the H-H interatomic distance, $d$. Obviously, 
the motion of the two atoms of a diatomic molecule is governed by six
coordinates, not just $Z$ and $d$, but the dependence
on the other (neglected) four coordinates was felt to be of minor
importance.
It was generally accepted that the PES in the considered two-dimensional
space (a so-called ``elbow plot'') should exhibit a strongly curved 
reaction path and a so-called late barrier towards dissociative adsorption, 
i.e. a barrier 
after the curved region of the PES, close to the surface, 
in order to account for strong vibrational effects in
dissociative adsorption and associative desorption.

Interestingly, strong vibrational heating  was also found for hydrogen
molecules desorbing from Pd\,(100)~\cite{Sch89}, although for this substrate,
in contrast to Cu, the dissociative adsorption is non-activated
\cite{Ren89,Sch92b,Wil95}. Nevertheless, within the spirit of the above
discussion
for the H$_2$/Cu system, Brenig {\em et al.}~\cite{Sch92,Bre94} 
reproduced the
vibrational heating for H$_2$ desorption from Pd\,(100) by quantum
dynamical calculations using two-dimensional model potentials with
a minimum barrier of 200~meV. Since this is in conflict with the
fact that adsorption of H$_2$ in non-activated, Darling and Holloway
\cite{Dar92b,Dar93} questioned the validity of this theoretical work
and argued that one has to take into account a distribution of
barrier heights 
which requires higher-dimensional calculations.
In their model calculations, which were still two-dimensional,
 they basically showed that it is not
possible to reproduce vibrational heating in desorption with a barrier-less
two-dimensional elbow potential \cite{Dar92b}.

Recently we enhanced the computational approach for
scattering of molecules at surfaces~\cite{Bre94,Bre93}
and
are now able to investigate reactions of diatomic 
molecules on surfaces with all six degrees of freedom of the molecule being 
treated quantum dynamically~\cite{Gro95}.
Such a study was performed for the sticking probability of
H$_2$ at Pd\,(100)~\cite{Gro95} employing
a high-dimensional PES derived from first-principles calculations~\cite{Wil95}.
In the present paper we use the same PES and the same six-dimensional (6-D)
quantum-dynamical method.
The PES for the interaction of H$_2$ and Pd\,(100) possesses non-activated
pathway, but also (in fact in majority) activated
pathways~\cite{Wil95}. It was found that H$_2$ molecules impinging
with small energies are efficiently steered along the
non-activated pathways towards dissociatice adsorption~\cite{Gro95}.
However, with increasing energy the steering effect gets less effective
and the majority of molecules proceeds via pathways with energy
barriers~\cite{Gro95}.
The steering mechanism is operative for the translational degree of freedom 
as well as for the rotations~\cite{Kay95}.
In the meantime two predictions of the quantum dynamical calculations for 
H$_2$/Pd\,(100), namely the strong rotational hindering of the dissociation
at low kinetic energy~\cite{Gro95,Gro96} and the orientation dependence
of adsorption and desorption~\cite{Gro95}, have been confirmed 
experimentally \cite{Beu95,Wet96}.

In order to investigate the effects of molecular vibrations
on the sticking probability we had to extend the calculations~\cite{Gro95}
to higher energies. 
To describe the dynamics of initially vibrating molecules
properly we have increased the number of vibrational eigenfunctions
in the expansion of the wave function of the hydrogen nuclei:
In the present study up to 25,200 channels per total energy are
taken into account compared to 21,000 channels which were considered
previously~\cite{Gro95}.
We will show that
in spite of the absence of a minimum barrier towards dissociative adsorption
and a strongly curved reaction path there are still substantial vibrational
effects in the adsorption and desorption of H$_2$/Pd\,(100). We will
demonstrate that they are caused by
a strong lowering of the H-H vibrational frequency
during the adsorption and by the multi-dimensionality of the PES relevant for
the dissociation process. However, some discrepancies to the experimental
results of refs.~\cite{Sch89,Sch92} remain.

\section{Results and Discussion}

\begin{figure}[tb]
\unitlength1cm
\begin{center}
   \begin{picture}(10,9.0)
      \includegraphics{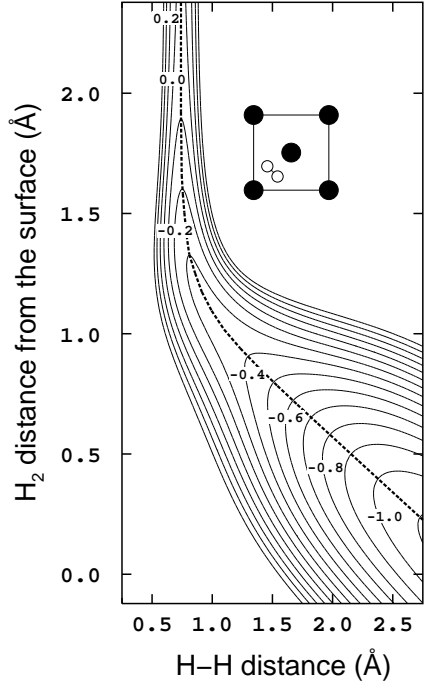}
   \end{picture}

\end{center}
   \caption{Contour plot of the PES along a
            two-dimensional cut through the
            six-dimensional coordinate space of H$_2$/Pd\,(100).
            The inset shows the
            orientation of the molecular axis and the lateral
            H$_2$ center-of-mass coordinates. The coordinates 
            in the figure are the H$_2$ center-of-mass distance 
            from the surface $Z$ and the H-H interatomic distance $d$. The 
            dashed line is the optimum reaction path.
            Energies are in eV per H$_2$ molecule.
            The contour spacing is 0.1~eV.  }

\label{elbow}
\end{figure}

Figure \ref{elbow} shows a cut through
our PES of H$_2$/Pd\,(100), where the most 
favorable path towards dissociative adsorption is marked by the dashed line. 
As discussed above, for this path there is no energy barrier hindering
dissociation, i.e., the adsorption is non-activated.
The curvature of the optimum reaction pathway (see dashed line in Fig.
\ref{elbow})
is relatively moderate compared to the curvature of reaction pathways
of previously assumed or guessed PESs which had been used
in earlier low-dimensional
studies~\cite{Jac87,Hal90,Kue91,Dar92a,Sch92,Bre94,Dar92b}.
The detailed total-energy calculations~\cite{Wil95} showed that the
PES is strongly anisotropic and corrugated so that besides non-activated
paths the majority of pathways towards dissociative adsorption
has in fact energy barriers with a rather broad
distribution of heights and positions.
The barrier height depends on the molecular orientation and impact site
in the surface unit cell. 

\begin{figure}[tb]
\unitlength1cm
\begin{center}
   \begin{picture}(10,6.1)
      \includegraphics{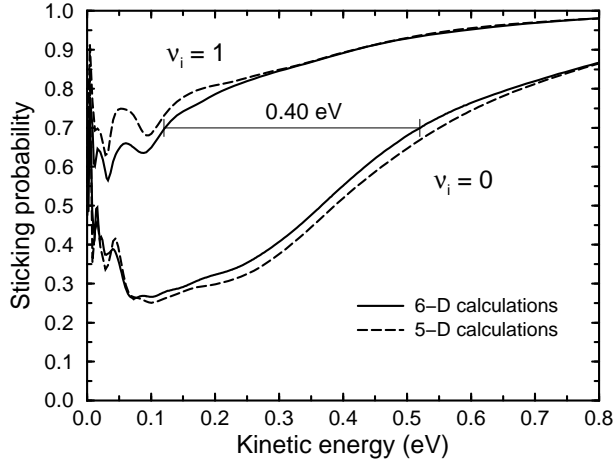}
   \end{picture}

\end{center}
   \caption{Sticking probability versus kinetic energy for
a H$_2$ beam under normal incidence on a Pd\,(100) surface.
The molecules are initially in the rotational
ground state $j_i = 0$, and in their initial  vibrational states
are $\nu_i =0$ (lower curves) and $\nu_i = 1$ (upper curves).
The solid lines show the results of the 6-D calculations and the dashed lines
are five-dimensional calculations where the  vibrational degree of freedom
is approximated by an adiabatic treatment (see text).}

\label{stickvib}
\end{figure}

Figure \ref{stickvib} presents results for the sticking probability as a 
function of the kinetic energy of a H$_2$ beam under normal incidence. 
Quantum mechanically determined sticking probabilities 
for hydrogen at surfaces with an attractive well exhibit an oscillatory
structure as a function of the incident energy~\cite{Gro95,Kay95,Dar90,Gro95b},
reflecting the opening
of new scattering channels and resonances \cite{Dar90,Gro95b}, as also
observed in, e.g, LEED \cite{LEED}.
Such oscillations have not been observed yet.
In the experiments the molecular beams are not strictly mono-energetic but
have a certain energetic broadening. For the calculations of
Fig.~\ref{stickvib} we assumed an energy width $\Delta E_{i}/E_i = 0.2$,
typical for experiments~\cite{Ren89}. As a consequence,
the quantum dynamical oscillations are smoothed out.

The solid curves correspond to 6-D calculations for
H$_2$ molecules initially in the vibrational ground and first excited
state, respectively. The $\nu_i = 0$ curve shows the characteristic
initial decrease  of the sticking probability with increasing energy
(for $E_{i} \lesssim 0.1$ eV) which is due to the decreasing importance
of the steering effect with increasing energy~\cite{Gro95}. Also for
molecules initially in the first
excited vibrational state a corresponding behavior is found, but for these
the steering effect is strong only for very small energies ($E_{i} \lesssim
50$ meV).
For kinetic energies higher than $\approx 50$~meV the $\nu_i = 1$ molecules
experience an increasing sticking probability
which is significantly larger than for  $\nu_i = 0$ molecules.

The effect of the initial vibrational state can be quantified by the
vibrational efficacy
\begin{equation}
\Xi_{\rm vib}(S ) \ = \ \frac{\epsilon_{\nu_i = 0} (S) - 
\epsilon_{\nu_i = 1} (S )}{\hbar \omega_{\rm vib}},
\end{equation}
where $\epsilon(S)$ is the kinetic energy required to obtain the
sticking probability $S$, so that $\Xi_{\rm vib}(S )$ is
the separation of the sticking curves for a certain sticking
probability divided by the gas-phase vibrational quantum
$\hbar \omega_{\rm vib} = 516$~meV. In Fig.~\ref{stickvib} 
we have marked $S = 0.7$, which leads to a value of the vibrational 
efficacy of $\Xi_{\rm vib} (S=0.7) = 0.75$. This means
that 75\% of the vibrational energy is effective in promoting the
dissociative adsorption, a value even higher than in the H$_2$/Cu(111)
system \cite{Ret95}.

\begin{figure}[tb]
\unitlength1cm
\begin{center}
   \begin{picture}(10,6.1)
      \includegraphics{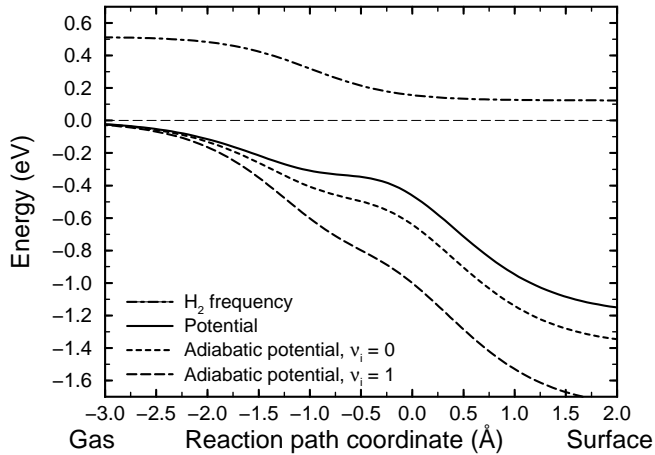}
   \end{picture}

\end{center}
   \caption{H-H vibrational frequency $\hbar \omega (s)$, 
           potential V$_0 (s)$ and ``vibrationally 
           adiabatic potentials'' (see Eq.~2) along the reaction path of 
           Fig.~\protect{\ref{elbow}}.
           $s = 0$ corresponds to the point of maximum curvature of the
           reaction path. }

\label{adiavib}
\end{figure}

To clarify why and how molecular vibrations
effect the sticking probability we performed calculations for a 
reduced coordinate space, namely allowing only for five
degrees of freedom for the two hydrogen atoms.
This reduction was achieved by keeping the molecules in their initial 
vibrational state. Although the vibrational state is kept fixed, the
energy of the vibration, which is determined by the strength of the
H-H interaction, will change along the scattering pathway.  
As can be seen in Fig.~\ref{stickvib},
these five-dimensional  results are very close to the 6-D results.
This reflects two facts. First, the molecular vibrational state is
a sufficiently good quantum number and is almost conserved during the 
scattering, i.e., the probability for transitions between different
vibrational states during the scattering event is rather low.
 And second,
the curvature of the reaction path of the H$_2$/Pd(100) PES is not crucial
for the vibrational effects in this system because in the 5-D calculations
no curvature is present in the Hamiltonian.

\begin{figure}[tb]
\unitlength1cm
\begin{center}
   \begin{picture}(10,6.1)
      \includegraphics{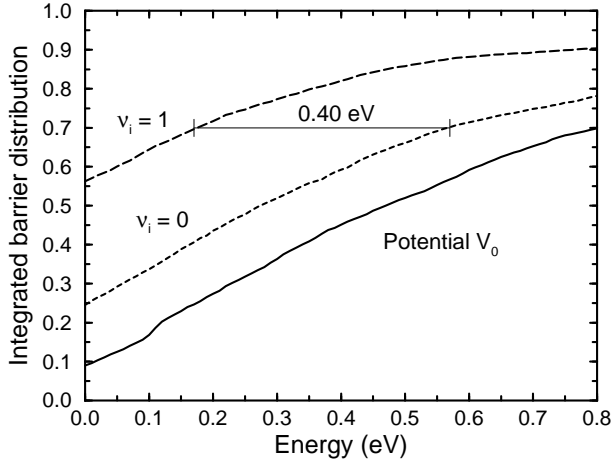}
   \end{picture}

\end{center}
   \caption{Integrated barrier distribution using the potential energy 
$V_{0}$  and ``vibrationally adiabatic potentials''.
}

\label{barrdist}
\end{figure}

As the next step we will analyze the
``vibrationally adiabatic potentials'' 
which are defined by
\begin{equation}
V_{\rm adia}^{\nu_i}(s) \ = \ V_0 (s) \ 
+ \ (\hbar \omega (s) - \hbar \omega_{\rm vib}) \ 
(\nu_i + \frac{1}{2} )\quad,
\end{equation}
where $s$ is the coordinate along the reaction path (see Fig.~\ref{elbow}).
The vibrationally adiabatic potential is the relevant potential
for the H$_2$ molecule moving on the PES in a particular 
vibrational state taking the change of the vibrational frequency
along the path into account.
In Fig.~\ref{adiavib} we have plotted the vibrational frequency
together with the potential and the vibrational adiabatic potentials for
$\nu_i = 0$ and $\nu_i = 1$ along the reaction path coordinate $s$ of 
Fig.~\ref{elbow}. At $s=0$, the point of maximum curvature in Fig.~\ref{elbow},
the vibrational frequency 
is strongly reduced from its gas phase value of 
$\hbar \omega = 516$~meV to about 150~meV. 
This leads to a lowering of the vibrationally adiabatic potential 
by 183~meV for $\nu_i =0$ and by 549~meV for $\nu_i =1$. 
Such a lowering does not only occur for the most favorable adsorption 
path, but also for other non-activated and activated pathways,
i.e., for other impact sites in the surface 
unit cell and for other molecular orientations. 
This can be demonstrated by the integrated barrier distribution for the
potential $V_0$ and the vibrationally adiabatic potentials
\begin{equation}\label{hole}
P_b (E) \ = \ \frac{1}{2 \pi A} \
\int \Theta (E - E_b (\theta,\phi,X,Y)) \cos\theta d\theta \ d\phi
\ dXdY.
\end{equation}
In Eq.~\ref{hole}, 
$\theta$ and $\phi$ are the polar and azimuthal orientation of the
molecule, and $X$ and $Y$ are the lateral coordinates of the
hydrogen center-of-mass. $A$ is the area of the surface unit cell.
Each quadruple defines a cut through the six-dimensional space (see
Fig.~\ref{elbow} for one example), and $E_b$ is the minimum energy barrier 
along such a cut. The function $\Theta$ is the Heavyside step function. 
The quantity
$P_b(E)$, which is plotted in Fig.~\ref{barrdist}, is the fraction 
of the configuration space, for which the barrier towards dissociative
adsorption is less than $E$; it corresponds to the sticking probability
in the classical sudden approximation, which forms the basic approximation
behind the so-called ``hole model'' \cite{Kar87}. Actually, the comparison
of Figs.~\ref{stickvib} and \ref{barrdist} reveals that the hole model 
gives a satisfactory description of the sticking probabilities at high kinetic
energies above 0.3~eV, whereas it is strongly at variance with full dynamical
calculations at low kinetic energies, where the steering effect is operative.

Figure~\ref{barrdist} demonstrates that the barrier distribution is 
lowered due to the decrease of the vibrational frequency
by about 180~meV for molecules in the vibrational ground state compared
to the potential $V_0(s)$ and by further $\approx$400~meV for 
molecules in the first excited vibrational state.
Because of the lowered potential the vibrationally excited molecules are
accelerated more strongly towards the surface, they become so fast that the 
steering mechanism is suppressed. For that reason the initial 
decrease of the sticking probability for $\nu_i =1$ is limited to
low energies ($E_{i} \lesssim 50$ meV, see Fig.~\ref{stickvib}).
The difference in the barrier heights of about 400~meV
is reflected by the energetic shift between the sticking curves
for $\nu_i = 0$ and $\nu_i =1$ in Fig.~\ref{stickvib}  
for sticking probabilities larger than $\approx$0.7.
Hence it is the strong decrease of the H-H vibrational frequency
during the dissociation which causes the vibrational effects in adsorption.

In figs. \ref{adiavib} and \ref{barrdist} zero-point effects due to
the frustrated rotation and translation parallel to the surface
are not taken into account since we like to concentrate here on states
with different vibrational quantum numbers. The influence of
frustrated rotation and translation of the H$_2$ molecule in contact
with the surface will be discussed in a forthcoming paper \cite{Gro96b}.

\begin{figure}[tb]
\unitlength1cm
\begin{center}
   \begin{picture}(10.,6.1)
      \includegraphics{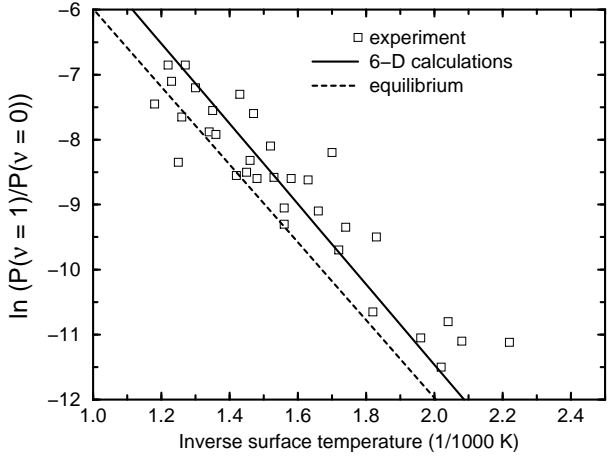}
   \end{picture}

\end{center}
   \caption{Vibrational excitation in desorption. 
       Boxes: experimental results (see text)
\protect{\cite{Sch92}}. 
              Dashed line: vibrational population in 
              thermal equilibrium with the surface temperature,
              solid line: 6-D calculations.}

\label{vibdes}
\end{figure}

According to
the principle of detailed balance, the strong enhancement
of the sticking probability of vibrationally excited molecules
implies a strong vibrational heating of molecules observed in associative
desorption.
Figure~\ref{vibdes} displays the logarithm of the population ratio of the
vibrationally first excited and the ground state in desorption versus the
inverse surface temperature. 
The theoretical values  were obtained by summing up
over all final rotational states.
The dashed line results from the assumption 
that the H$_2$ vibrations are in thermal equilibrium
with the surface temperature. Indeed we find vibrational
heating in our calculations. At $T_s = 700$~K the ratio is
2.5 times higher than for thermal equilibrium.
Absolute values of the vibrational heating in hydrogen desorption 
from Pd(100) were only measured for D$_2$, not for H$_2$.
In the system H$_2$/Ni(110), which has a similar sticking curve as
H$_2$/Pd\,(100) \cite{Ren89}, a vibrational heating of a factor of two
was found \cite{Win94}, in good agreement with our results.
The experimental results in Fig.~\ref{vibdes} are only determined
within a calibration factor~ \cite{Sch92}. This means that only
the slope of the experimental data has significance,
not the absolute values. From this slope an apparent activation 
energy $E_a = 428 \pm 30$~meV \cite{Sch92} has been deduced. Our theoretical
work gives  $E_a = 519 \pm 1$~meV.
Considering the scatter in the experimental data~(see Fig.~\ref{vibdes})
and the fact that the theory does not employ any empirical parameters,
the comparison of the experimentally and theoretically obtained
apparent barrier is satisfactory.

In fact, an analysis of the measured sticking 
probability of H$_2$/Pd(100) \cite{Ren89} makes it plausible 
that the apparent activation energy should be close
to the gas-phase frequency of H$_2$.
If we restrict ourselves for the sake of clarity to the two
coordinates $Z$ and $d$, then according to the principle of detailed balance 
the vibrational population ratio in desorption is given by
\begin{equation}\label{vibpop}
\frac{P_1}{P_0} \ = \ \frac{\int \ e^{-E/k_B T_s} \ S(E,\nu_i = 1) \ dE}
{\int \  e^{-E/k_B T_s} \ S(E,\nu_i = 0) \ dE} \ \cdot 
\exp \left( \frac{-\hbar \omega_{\rm vib}}{k_B T_s} \right),
\end{equation}
where $S(E,\nu_i)$ is the sticking probability for initial vibrational state 
$\nu_i$ and kinetic energy $E$. Now an analysis of Eq.~\ref{vibpop} yields
that a strong lowering of the apparent activation energy from the gas-phase 
vibrational energy can only be caused by a ratio 
\mbox{$S(E,\nu_i=1)$/$S(E,\nu_i=0)$} 
which decreases exponentially with increasing
kinetic energy. This would only be the case if \mbox{$S(E,\nu_i=0)$} increased 
exponentially with increasing energy,
which requires the assumption of a minimum barrier towards
dissociative adsorption, as was done in the calculations of 
refs.~\cite{Sch92,Bre94}. However, the sticking in the system H$_2$/Pd\,(100) 
is non-activated and initially decreases with increasing 
kinetic energy \cite{Ren89}.
Such a sticking probability should not cause a lowering of the apparent 
activation energy for vibrational excitation in desorption, as reproduced 
by our calculations. Either the principle of detailed balance is not
directly applicable for the adsorption/desorption of H$_2$/\,Pd(100), which 
seems to be improbable from the experience of hydrogen-on-metal systems 
\cite{Dar95}, or there is an inconsistency between the adsorption and 
desorption experiments.

We further note that
the apparent activation energy in our calculations
depends on the rotational states of the molecules. If one considers, e.g.,
only $j_f = 4$ rotational states, where $j_f$ is the 
final rotational quantum number, for the vibrational excitation in desorption,
then the theoretical activation energy drops to $E_a = 487 \pm 2$~meV.  
This is due to the fact that the sticking probability for molecules in
the vibrational ground state becomes small (less than 0.1) for
$j_i \geq 4$ \cite{Gro96}, while for vibrationally excited molecules the
sticking probability is almost independent of the initial rotational state. 
Also the absolute value of the vibrational heating depends 
sensitively on the rotational quantum number; 
for $j_f =4$ the vibrational heating rises to a factor of 3.6.

For D$_2$/Pd\,(100) a vibrational over-population in desorption 
of $\nu =1$ by a factor 
of nine was found \cite{Sch89}, which is much higher than our result for
H$_2$/Pd\,(100). 
Note that the interaction of D$_2$ with Pd\,(100) is given by {\em exactly}
the same PES as for H$_2$. At present, full quantum dynamical calculations
for D$_2$ are not feasible because the number of relevant, energetically
accessible channels is significantly higher than for H$_2$ due to the
larger mass and the therefore smaller energetic level spacings. We do not
expect, however, that the calculations for D$_2$ would yield a vibrational
heating of a factor of nine. Again analyzing Eq.~\ref{vibpop}, for 
such a large vibrational heating the sticking probability for $\nu_i = 0$ 
states of D$_2$ has to be below 0.1 for all energies, 
much lower than for H$_2$, but already the
hole model, that does not take into account the steering effect, yields
higher values (see Fig.~\ref{barrdist}).

The reason for this discrepancy may lie in the determination of the PES.
On the other hand, if all experimental results were correct,
the application of the principle of detailed balance 
for the adsorption/desorption of hydrogen/Pd\,(100)
would yield a large isotope effect, which, 
for example, has not been found for hydrogen on Pd\,(111) \cite{Beu95}.
It may also be that our understanding of the adsorption/desorption
dynamics is still incomplete.

\section{Conclusions}

In conclusion, we reported a six-dimensional quantum dynamical
study of dissociative adsorption on and associative desorption from
H$_2$/Pd\,(100).
 We have shown that large vibrational effects
in dissociative adsorption and associative desorption of H$_2$/Pd\,(100)
exist. They are not due to a strongly curved reaction path and a late 
minimum barrier to adsorption,
as was previously assumed, but they are caused
by the strong lowering of the H-H vibrational 
frequency during the adsorption and the multi-dimensionality of the
relevant phase space with its broad distribution of barrier
heights. Quantitative differences between experiment and theory
and inconsistencies between adsorption and desorption experiments
are identified which deserve further clarification.

\end{document}